\documentclass[]{emulateapj}
\usepackage{graphicx}
\usepackage{color}

\shorttitle{The disruption of cusps}
\shortauthors{Tobias Goerdt et al.}

\begin{document}

\title{Core creation in galaxies and haloes via sinking massive objects}

\author{Tobias Goerdt\altaffilmark{1}*, Ben Moore\altaffilmark{2}, J. I.
Read\altaffilmark{3,2} and Joachim Stadel\altaffilmark{2}}
\affil{$^1$Racah Institute of Physics, The Hebrew University, Jerusalem 91904,
Israel\\$^2$Institut f\"ur Theoretische Physik, Universit\"at Z\"urich,
Winterthurerstrasse 190, CH-8057 Z\"urich, Schweiz \\$^3$Department of Physics
\& Astronomy, University of Leicester,
University Road, Leicester, LE1 7RH, UK}
\email{* tobias.goerdt@uam.es}

\begin{abstract}
We perform a detailed investigation into the disruption of central cusps via
the transfer of energy from sinking massive objects. Constant density inner
regions form at the radius where the enclosed mass approximately matches the
mass of the infalling body. We explore parameter space using numerical
simulations and give an empirical relation for the size of the resulting core
within structures that have different initial cusp slopes. We find that
infalling bodies always stall at the edge of these newly formed cores,
experiencing no dynamical friction over many dynamical times. As applications,
we consider the resulting decrease in the dark matter annihilation flux due to
centrally destroyed cusps, and we present a new theory for the formation of
close binary nuclei -- the `stalled binary' model. We focus on one particularly
interesting binary nucleus system, the dwarf spheroidal galaxy VCC~128 which is
dark matter dominated at all radii. We show that its nuclei would rapidly
coalesce within a few million years if it has a central dark matter cusp slope
steeper than $r^{-1}$. However, if its initial dark matter cusp is slightly
shallower than a log slope of $-0.75$ at $\sim 0.1$\% of the virial radius,
then the sinking nuclei naturally create a core equal to their observed
separation and stall. This is close to the log slope measured a recent billion
particle CDM halo simulation.
\end{abstract}

\keywords{cosmology: theory -- dark matter -- galaxies: dwarf -- 
galaxies: individual: VCC~128 -- methods: numerical}

\section{Introduction}\label{sec:intro}
Massive objects orbiting within a cuspy mass distribution are expected to lose
momentum and sink via dynamical friction \citep{chandrasekhar, white,
hernquist, capuzzo}. While spiralling inwards the massive perturber transfers
momentum to central particles/stars etc, moving them to a larger orbital
radius. One effect of this process is to make an initially cuspy dark matter
distribution shallower \citep{amr, amr2, tonini, ma, jardel}.

In the prevailing $\Lambda$CDM cosmology, dark matter haloes are cuspy, having
an inner density slope $\rho(r) \propto r^{-\gamma}$ with $\gamma > 1$ beyond
$\approx$ 1\% of the virial radius \citep[e.g.][]{dubinski, navarro, diemand}.
By contrast, observations of dwarf galaxies seem to indicate that they have a
cored mass distribution \citep[e.g.][]{deblok, sanchez, kleyna}, while
controversial evidence for cored mass distributions in dwarf spiral galaxies
has been debated for over a decade \citep{moore2}.

Previously published work has demonstrated numerically \citep{amr, amr2,
romano, ma, merritt, justin, jardel} and semi-analytically \citep{tonini} that
a sinking massive compact object -- a {\it perturber} -- will transfer energy
and angular momentum to the background via dynamical friction, creating a
central constant density core from an initially cuspy density distribution.
Once a core has formed, dynamical friction is no longer effective \citep{mich,
justin, inoue}. Dynamical arguments show that sinking perturbers will stall at
the outer edge of a core \citep{justin,inoue, inoue2}.

Here we quantify this stalling behaviour as a function of perturber mass
$M_{\rm pert}$ and central cusp slope $\gamma$. We consider a much larger range
in $M_{\rm pert}$ and $\gamma$ than in previous papers \citep{mich, justin} and
find that stalling persists even at very high perturber mass, as found also
recently by \citet{gualandris}. We also investigate how the cusp is physically
transformed into a core. 

The core formation mechanism we study is just one of several ways in which
central cores can be formed. Core formation can also proceed by three-body
encounters with a supermassive black hole binary \citep{milosavljevic}, as a
result of rapid mass loss due to supernovae outflows \citep[e.g.][]{navarro2,
justin3}, or as a result of the rapid ejection of a central supermassive black
hole due to anisotropic gravitational radiation recoil
\citep[e.g.][]{boylankolchin}. If these mechanisms play an important role then
our derived stalling radii as a function of $M_{\rm pert}$ and $\gamma$ will be
lower bounds.

We consider two applications of our results. The first is the effect of
cusp-destruction on the expected dark matter annihilation signal from galaxies
and dwarf galaxies. For a wide range of popular dark matter particle models,
dark matter can self-annihilate to produce $\gamma$-rays \citep{gunn}. Since
the signal goes as the dark matter density squared it is sensitive to the
central density distribution \citep{silk, lake}. For our second application, we
present a new theory for the formation of close binary nuclei -- the `stalled
binary' model. Close binary nuclei with projected separation $< 100$\,pc have
been observed on a range of scales in the Universe \citep[e.g.][]{lauer1,
lauer, lauer2, bender, houghton, mast, victor}. The standard model for these
has become the \cite{tremaine} eccentric disc model originally proposed to
explain M31 \citep{lauer2}. However, when the double nucleus of M31 was first
discovered, \cite{lauer1} speculated that the two bright nuclei really were
just that -- one from M31 and the other the cannibalised centre of a smaller
merged galaxy. The primary argument against this was that dynamical friction
would cause the nuclei to rapidly coalesce. We show that, as binary nuclei sink
via dynamical friction, they create a central constant density core. They then
stall at the edge of this core experiencing no further friction over many
dynamical times. We apply our new `stalled binary' model to one particularly
interesting binary nucleus system -- the dwarf spheroidal galaxy VCC~128
discovered by \citet{victor}. We show that this galaxy is dark matter dominated
at all radii. As a result -- if its binary nucleus is a `stalled binary' --
VCC~128 gives us a unique opportunity to constrain the central log-slope of the
dark matter density profile on very small scales. 

This paper is organised as follows: In \S2 we describe our analytical
framework, which is supported using $N$-body simulations. In \S3 we apply our
findings to dark matter annihilation, to binary nuclei, and to the special case
of VCC~128. Finally, in \S4 we present our conclusions.

\section{Transforming cusps to cores}
\label{sec:trans}
\subsection{Numerical models}

We use the split-power law $\alpha, \beta, \gamma$ model for our initial
background distribution \citep{abc,saha,zhao}:
\begin{equation}
\rho(r)=\frac{\rho_0} {\left({r /r_{\rm s}}\right)^\gamma \left[{1 +
\left({r/r_{\rm s}}\right)^{\alpha}}\right]^{\left(\beta - \gamma\right)/
{\alpha}}} \qquad (r \le R_{\rm vir})
\label{eq:den}
\end{equation}
where $\rho_0$ and $r_{\rm s}$ are the normalisation of the density and scale
length respectively, $\gamma$ is the inner log slope, $\beta$ is the outer log
slope, whereas $\alpha$ controls the transition between the inner and the
outer region. Since we focus on the very inner regions of the halo in our
analysis, $\alpha$ and $\beta$ are not critical. We fix $\alpha = 1$ and
$\beta=3$ -- the commonly accepted values for cold dark matter haloes. We
explore a range of values for the central cusp slope $\gamma$.

\begin{table}
\begin{center}
\setlength{\arrayrulewidth}{0.5mm}
\begin{tabular}{cccccc}
\hline
{\it Halo} & $\gamma$ & $\rho_0$ / M$_\odot$pc$^{-3}$ & $r_{\rm s}$ / kpc & con
& M$_{\rm vir}$ / M$_\odot$ \\
\hline
A & 1.75 & 0.000232 & 20.8 & 3.88 & $3.10 \times 10^{10}$\\
B & 1.50 & 0.001333 & 10.3 & 7.50 & $2.71 \times 10^{10}$\\
C & 1.25 & 0.004029 & 6.59 & 11.5 & $2.57 \times 10^{10}$\\
D & 1.00 & 0.009109 & 5.01 & 15.0 & $2.50 \times 10^{10}$\\
E & 0.75 & 0.017732 & 3.91 & 19.0 & $2.43 \times 10^{10}$\\
F & 0.50 & 0.027746 & 3.38 & 22.0 & $2.40 \times 10^{10}$\\
\hline
\end{tabular}
\end{center}
\caption{Parameters for the six different dark matter haloes we use in
our analytical and numerical calculations.}
\label{tab:haloes}
\end{table}

\begin{figure*}
\begin{center}
\includegraphics[width=5.95cm]{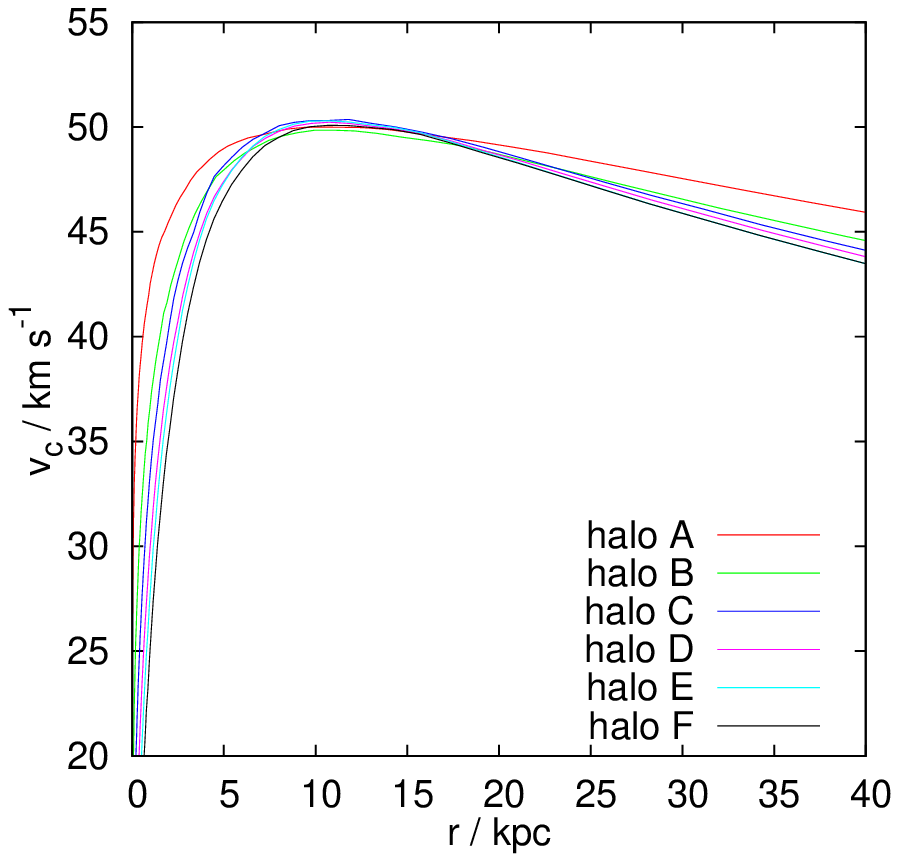}
\includegraphics[width=5.95cm]{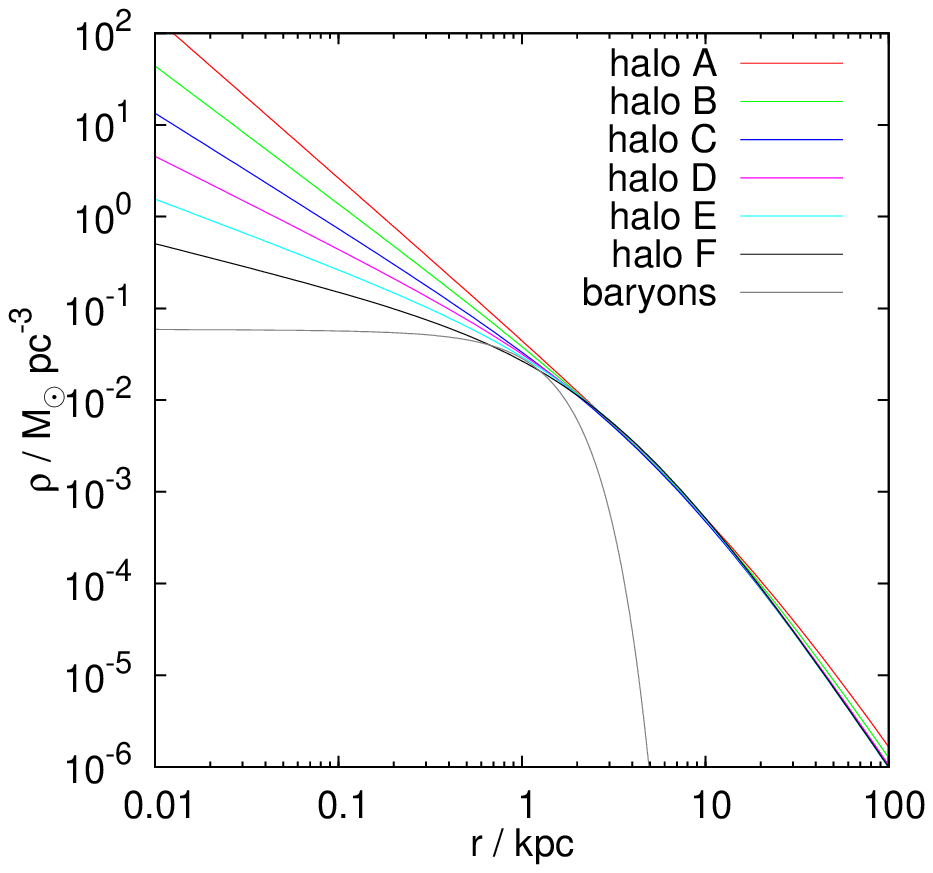}
\includegraphics[width=5.95cm]{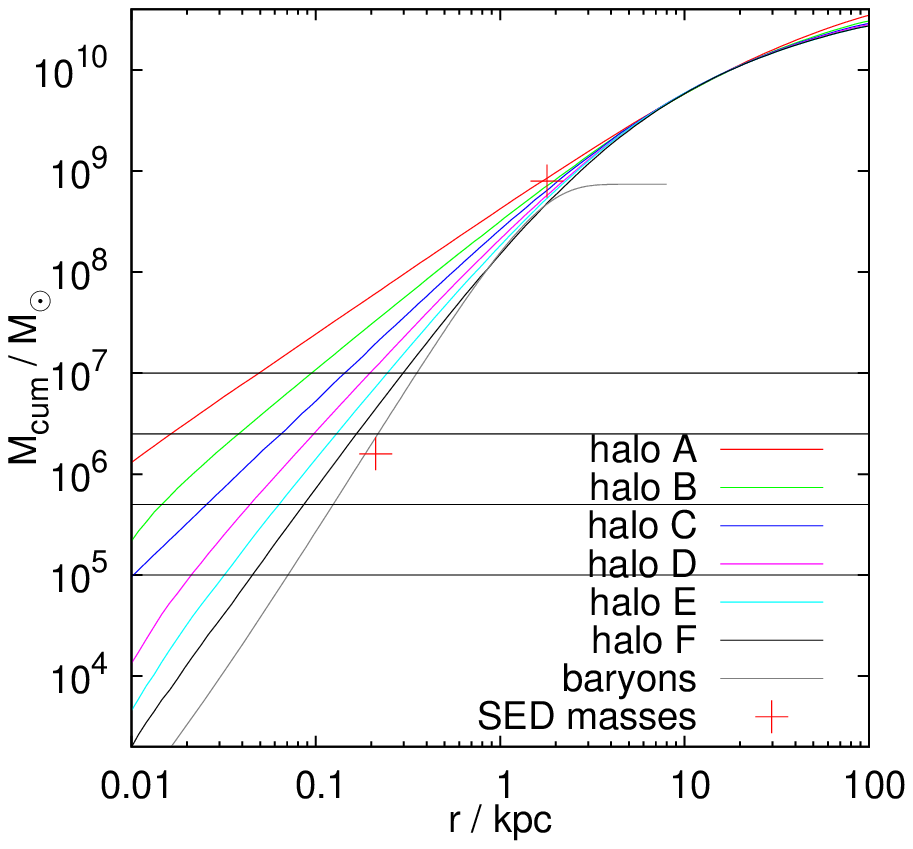}
\end{center}
\caption{Circular velocities ({\it left}) radial density profiles ({\it
middle}) and cumulative mass ({\it right}) for all our haloes A -- F. These are
constructed to have peak values of 50 km s$^{-1}$ at 10.75 kpc. The horizontal
black lines on the right panel show the perturber masses $M_{\rm pert}$ we use
in the simulations (\S \ref{sec:nbody}). The grey lines and the red crosses show
estimates of the baryonic (stellar) mass distribution of the dSph galaxy
VCC~128. Notice that in all models, VCC~128 is dark matter dominated at all
radii.}
\label{fig:f1}
\end{figure*}

We will take our fiducial model to represent a low mass dark matter halo
typical of those surrounding dwarf galaxies with a maximum circular velocity
$v_{\rm peak}$ = 50 km s$^{-1}$. As our application later, we will consider the
Virgo cluster dwarf galaxy VCC~128 which has an absolute bolometric luminosity
of M$_{\rm B} = -15.5$\,mag \citep{victor}. We estimate its maximum circular
velocity to be $v_{\rm peak}$ = 35 - 65 km s$^{-1}$ using the Faber-Jackson
relation of dEs \citep{derijcke}. We adopt a concentration parameter of 15 for
$\gamma = 1.0$, which is a typical value found for cosmologically motivated
dwarf spheroidals \citep{lokas2}. This gives a $r(v_{\rm peak})$ of 10.75\,kpc.
For our other models we keep $v_{\rm peak}$ as well as $r(v_{\rm peak})$ constant
and only vary $\gamma$. This leads to the parameters given in Table
\ref{tab:haloes} and to the circular velocity curves and radial density
profiles, which are shown in Figure \ref{fig:f1}.

The dynamical friction sinking timescale of a perturber in such a system can be
derived using the Chandrasekhar dynamical friction formula
\citep{chandrasekhar}. Assuming that the perturber is always on a circular
orbit, and that the background particles have a Maxwellian velocity
distribution, we can calculate the infall rate:
\begin{eqnarray}
{{\rm d}r \over {\rm d}t} & = & -{4 \pi {\rm ln} \Lambda (r) \rho(r) G^2 M_{\rm
pert} r \over v^2_{\rm c}(r) {\rm d}\left[rv_{\rm c}(r)\right] / {\rm d}r}
\left\{{\rm erf} \left[{v_{\rm c}(r) \over \sqrt{2} \sigma(r)}\right]\right.
\nonumber \\ & - &
\left.{2 v_{\rm c}(r) \over \sqrt{2 \pi} \sigma(r)}{\rm exp} \left[{-v_{\rm
c}^2(r) \over 2 \sigma^2(r)}\right]\right\},
\label{eq:L1}
\end{eqnarray}
where $v_{\rm c}(r)$ is the circular speed at radius $r$; $M_\mathrm{pert}$ is
the mass of the in-falling body; $\ln\Lambda(r)$ is the Coulomb logarithm
$[\Lambda = b_{\rm max} / b_{\rm min}]$; $\rho(r)$ is the density of the dark
matter halo at radius $r$ according to equation (\ref{eq:den}); and $\sigma
(r)$ is the one-dimensional velocity dispersion of the halo. 

Equation (\ref{eq:L1}) assumes that dynamical friction is a local process that
proceeds by momentum exchange between the perturber and the background. This
cannot be the whole story. Firstly, it implies that if $\rho(r)\rightarrow 0$
there will be no dynamical friction. Secondly, it implies that if $\rho(r)
\rightarrow \mathrm{const}$, nothing special should happen and dynamical
friction should proceed as for any other $\rho(r)$. Both of these statements
conflict with numerical results \citep{Lin83, Trem84, Colpi, mich, justin}. The
implication is that dynamical friction is not only a local process, but also
depends on global resonances \citep{Trem84, sellwood, justin}. In the former
case, resonances (and therefore friction) are present even outside of a galaxy
where $\rho(r)=0$. In the latter case, $\rho = \mathrm{const.}$ is a harmonic
potential which is especially resonant. After a period of enhanced friction,
the perturber and the background reach a stable state with no net momentum
exchange \citep{kalnajs, justin}. Despite the above difficulties, in most
situations, equation (\ref{eq:L1}) works remarkably well. 

\subsection{Results from N-body simulations}
\label{sec:nbody}

We now study the response of different central cusps to sinking perturbers with
a range of masses using $N$-body simulations. We consider all of our haloes A
-- F presented in Table \ref{tab:haloes}. Equilibrium $N$-body representations
of these haloes were created with the algorithm described in \citet{stelios2}
using the multimass technique of \citet{marcel}. This gives structures that
have an effective resolution in the region of interest equivalent to using
$\sim 10^{10}$ single mass particles. The softening of the lightest particles
is 0.1 pc. We ran a grid of simulations with $M_{\rm pert} = [10^5,\,5\times
10^5,\, 2.5\times 10^6,\, 10^7,\, 5\times 10^7]\,$M$_\odot$ for each of the
haloes A, B, C, D, E, F corresponding to varying the initial central log
density slope: $\gamma = [1.75,\,1.5,\,1.25,\,1,\, 0.75,\,0.5]$ (see Table
\ref{tab:haloes}). The perturbers have softening lengths of 2 pc and were
started at a radius of 0.4~kpc (within the cusp region), except for the heaviest
perturbers which we had to start further out for reasons which will be
explained in \S\ref{sec:kick}. (We could have started all nuclei at the
outermost radius and obtained the same results, but this would have produced
avoidable computational costs. See Figure \ref{fig:f20}.) All simulations are
shown using circular orbits but similar scaling laws were found using more
eccentric orbits.

\begin{figure*}
\begin{center}
\includegraphics[width=5.95cm]{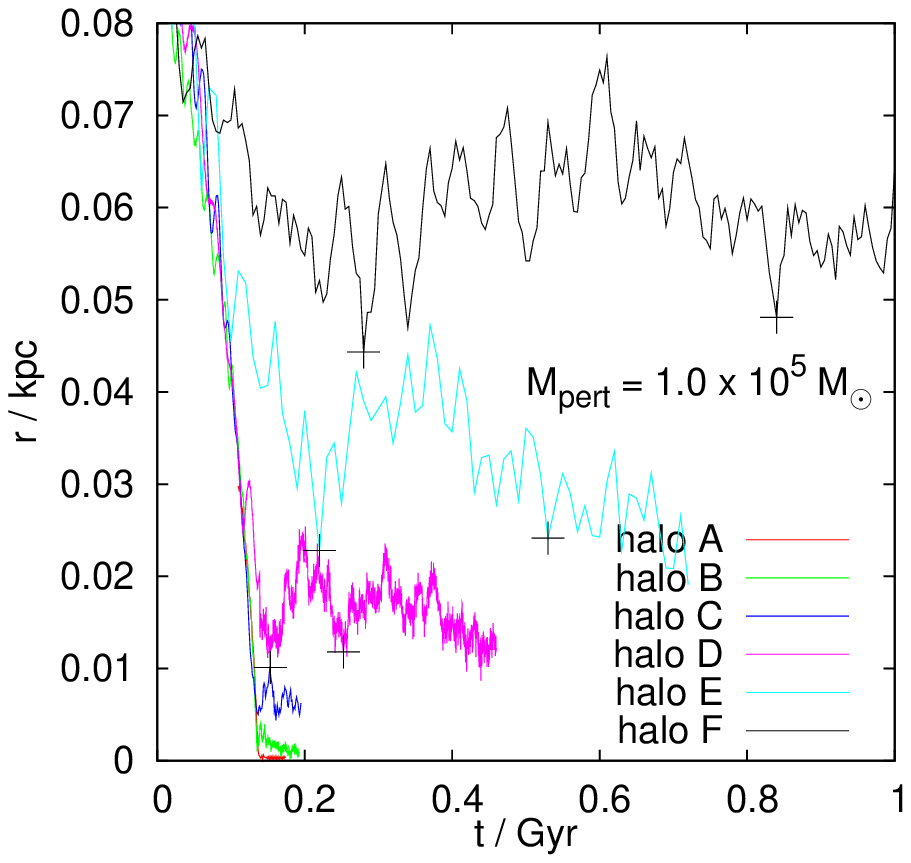}
\includegraphics[width=5.95cm]{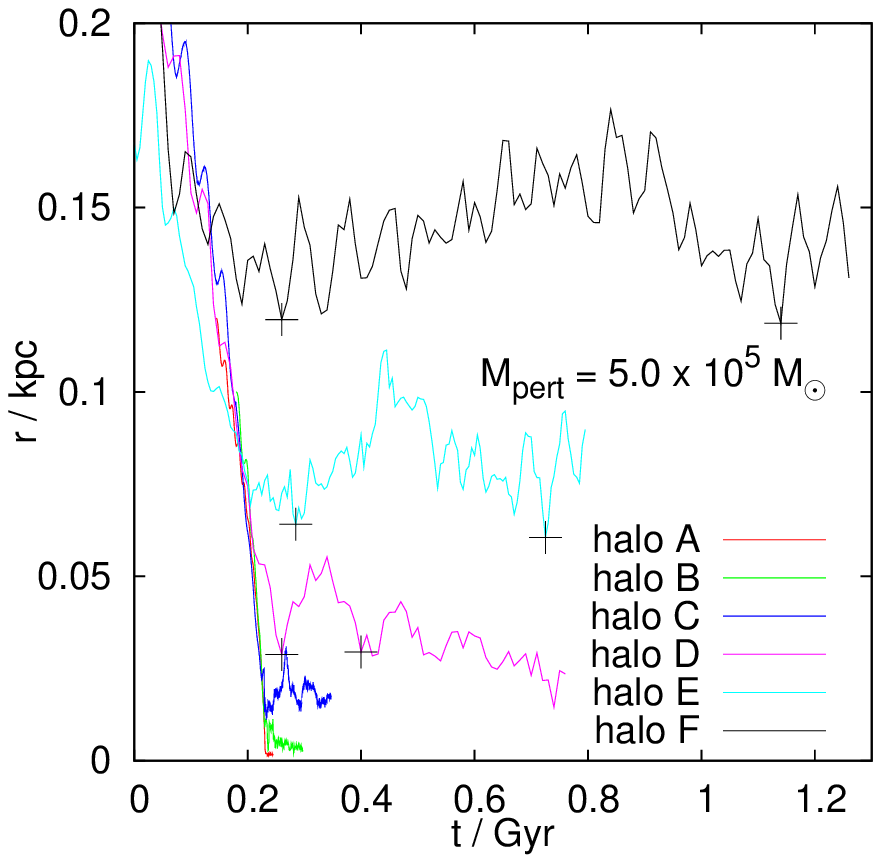}
\includegraphics[width=5.95cm]{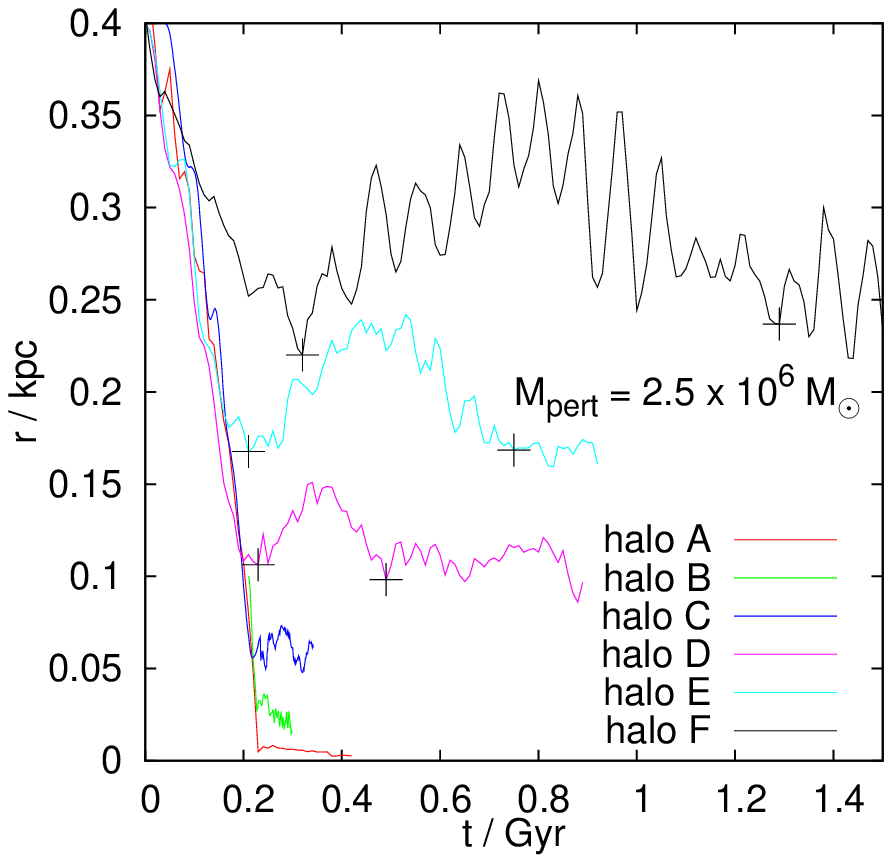}\\
\includegraphics[width=5.95cm]{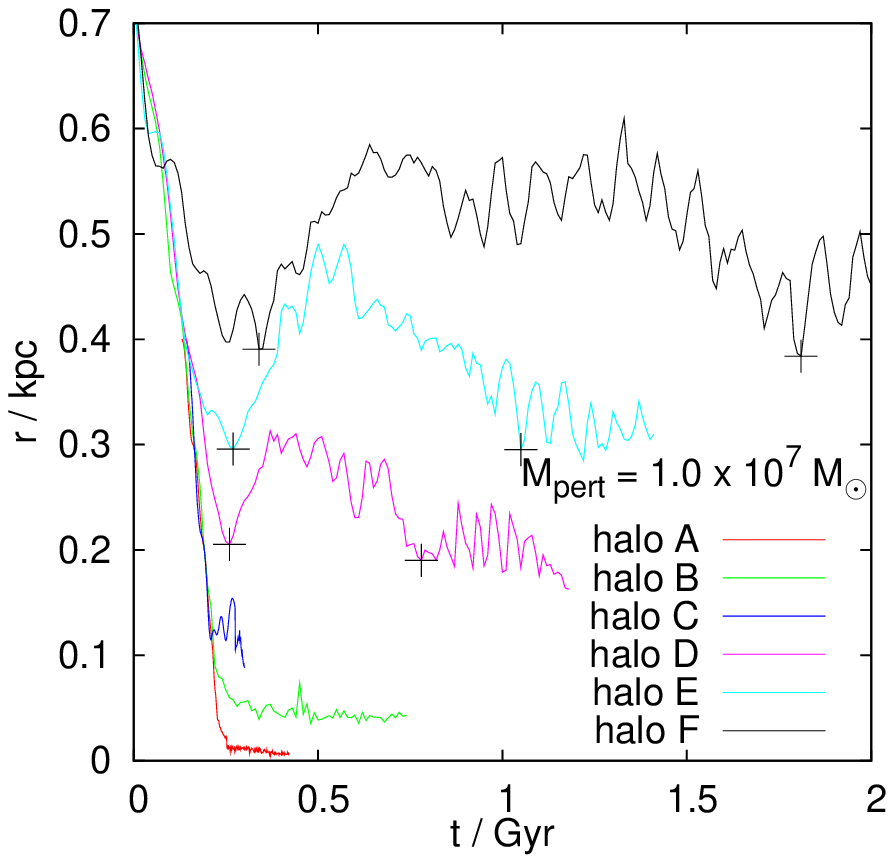}
\includegraphics[width=5.95cm]{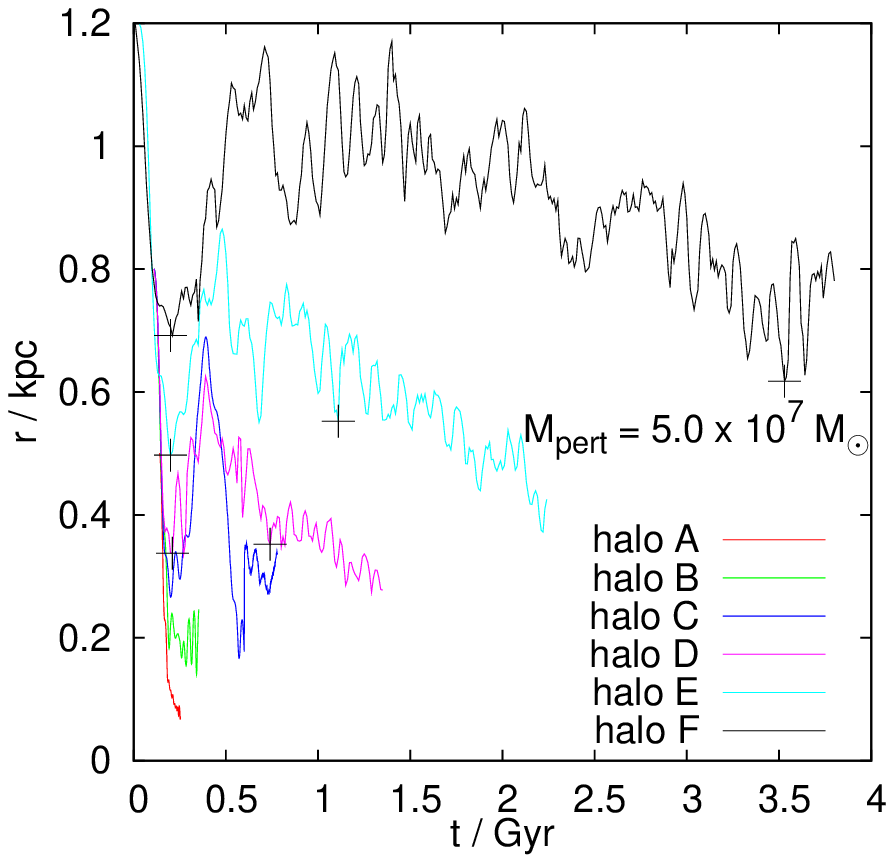}
\end{center}
\caption{Simulated position of a single perturber within different haloes. From
left to right the upper panels show perturber mass $M_\mathrm{pert} = [10^5,\,5
\times10^5,\,2.5\times 10^6]\,$M$_\odot$ and the lower panels mass
$M_\mathrm{pert} =[10^7,\,5 \times 10^7]\,$M$_\odot$. The black crosses
indicate the first point of closest approach (fpca) and the second point of
closest approach (spca).}
\label{fig:rgc0}
\end{figure*}

It is well known that if you pick the wrong centre, a cuspy profile will appear
to have a core \citep{beers}. We used the particle with the lowest potential as
the centre of the halo. In order to have a robust determination of the centre,
we removed the perturber and only used the lightest halo particles. In
addition, we double-checked our results against the `shrinking spheres method'
\citep[e.g.][]{power, mich, justin2}. The results agreed up to the noise level.
The resulting trajectories are given in Figure \ref{fig:rgc0}.

\begin{figure}
\includegraphics[width=8.63cm]{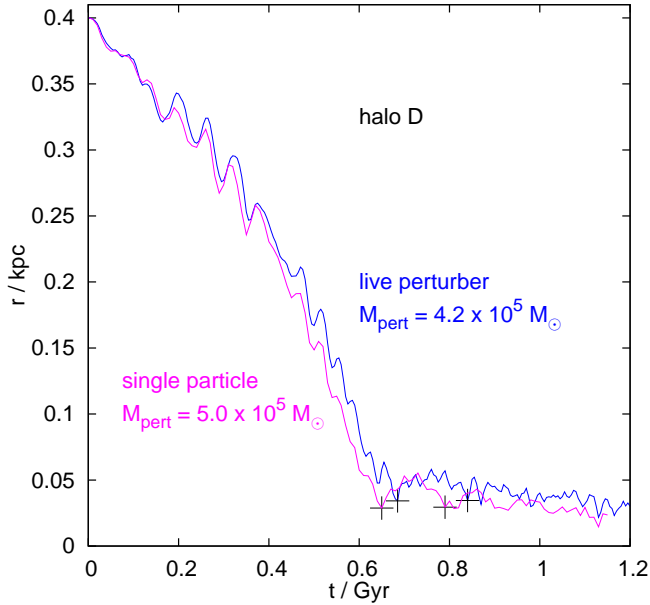}
\caption{Simulated position of a live perturber of mass $M_{\rm pert} = 4.2
\times 10^5$\,M$_{\odot}$ compared to a single particle perturber of mass
$M_{\rm pert} = 5.0 \times 10^5$\,M$_{\odot}$ both within halo D. The black
crosses indicate the respective first point of closest approach (fpca) and
second point of closest approach (spca).}
\label{fig:live}
\end{figure}

In order to double check the validity of modelling the perturber as just one
single particle, we a run an additional simulation with a live `globular
cluster' as perturber. The live globular cluster we use consists of $10^5$
particles which are distributed according to the King model \citep{king, michie,
bodenheimer} given by
\begin{eqnarray}
\rho(\Psi) & = & \rho_1 \exp \left({\Psi \over \sigma^2} \right) {\rm erf}
\left({\sqrt{\Psi} \over \sigma} \right) \nonumber \\ & - & \rho_1 \sqrt{4\Psi
\over \pi \sigma^2} \left(1+{2 \Psi \over 3 \sigma^2} \right),
\end{eqnarray}
where $\sigma$ is the velocity dispersion, $\rho$ is the density and $\Psi$ is
the relative potential. Each globular cluster is constructed with a W$_0 = 
\Psi(0) / \sigma^2$ parameter of 6, a total mass of $4.2 \times 10^5$
M$_{\odot}$, and a central velocity dispersion of 11\,km/s. We use 0.05\,pc for
the gravitational softening lengths of its particles. This perturber is put
into halo D at an initial distance of 0.4 kpc. Its trajectory can be seen in
Figure \ref{fig:live}. As one can clearly see  the behaviour of the live
perturber matches the behaviour of the single particle in this figure as well
as in Figure \ref{fig:rgc0} very well.

\subsubsection{Cusp destruction, core creation and stalling}
\label{sec:kick}

For all trajectories there is an apparent ``kickback" which occurs after a first
point of closest approach (fpca). The perturber seems to move away for a while,
reaches a maximum, and then returns to a second point of closest approach
(spca), where it finally stalls. For especially pronounced kickbacks, fpca and
spca are marked by black crosses in Figure \ref{fig:rgc0}. This apparent
``kickback" occurs at a point where the acceleration on the perturber due to
the background is equal to the acceleration on the background due to the
perturber, and the centre of mass of the system is significantly displaced. At
this point the true `centre' of the system becomes poorly defined. For this
reason, the ``kickback" feature is not physical but rather an artifact of our
centring algorithm. After the ``kickback",  the background rapidly rearranges
itself to form a central constant density core, at which point the perturber
stalls.

\begin{figure*}
\begin{center}
\includegraphics[width=5.95cm]{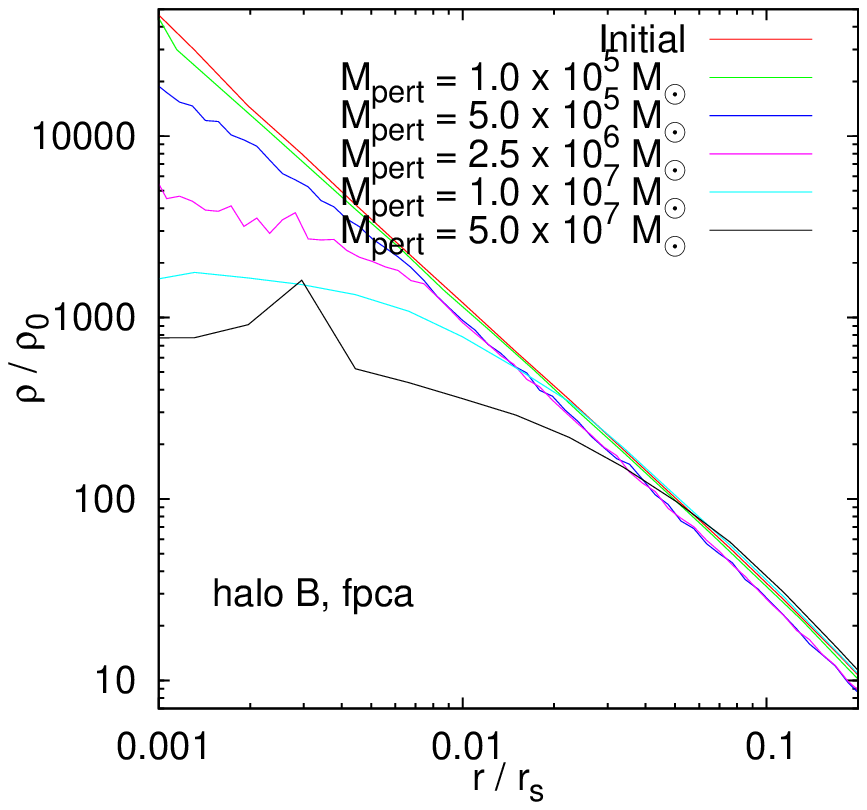}
\includegraphics[width=5.95cm]{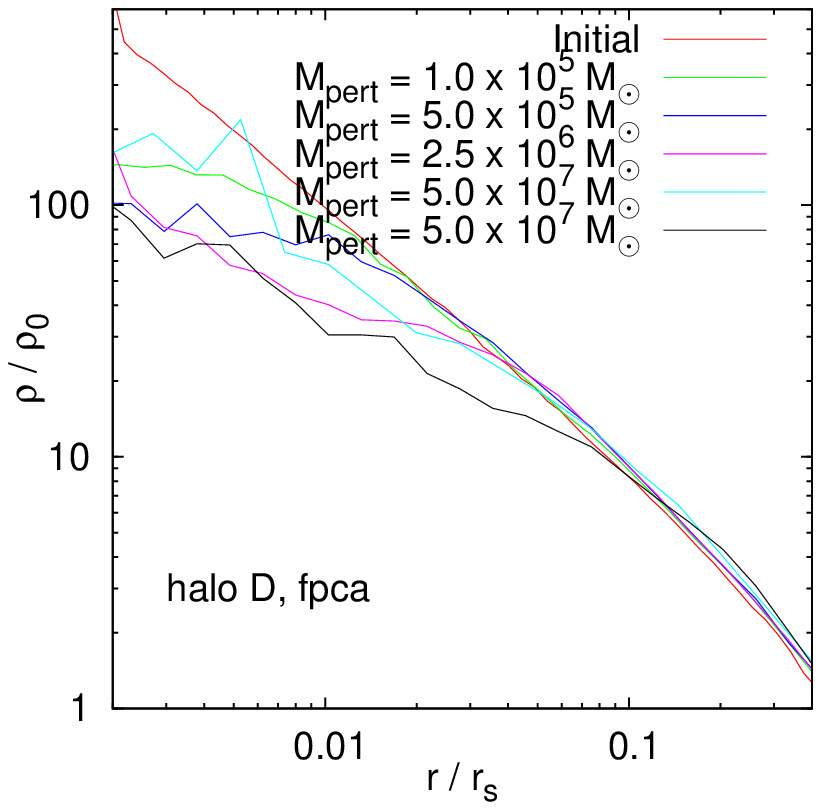}
\includegraphics[width=5.95cm]{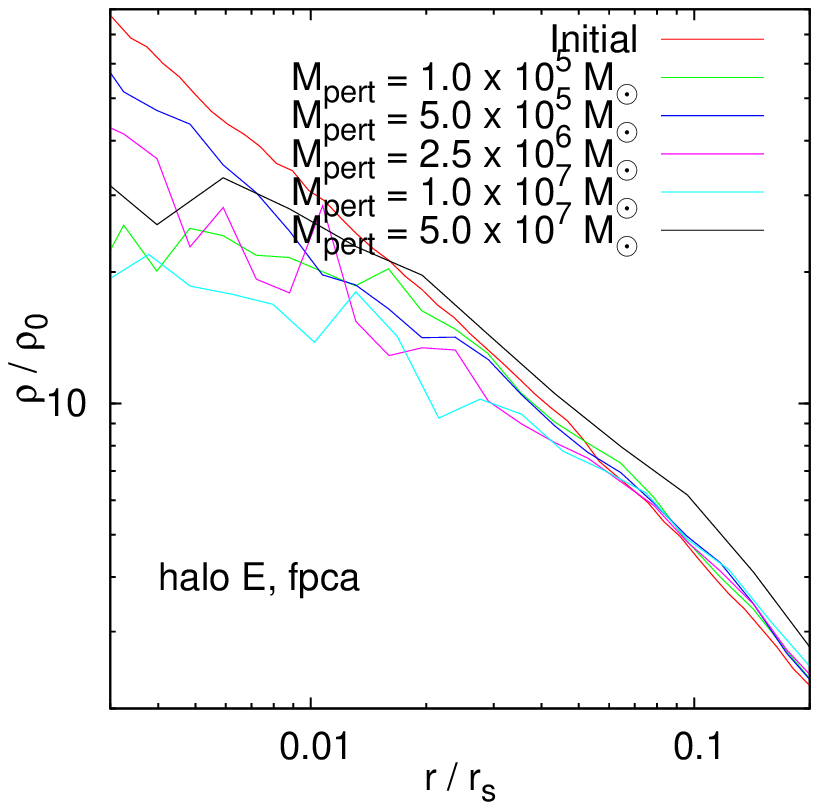}
\includegraphics[width=5.95cm]{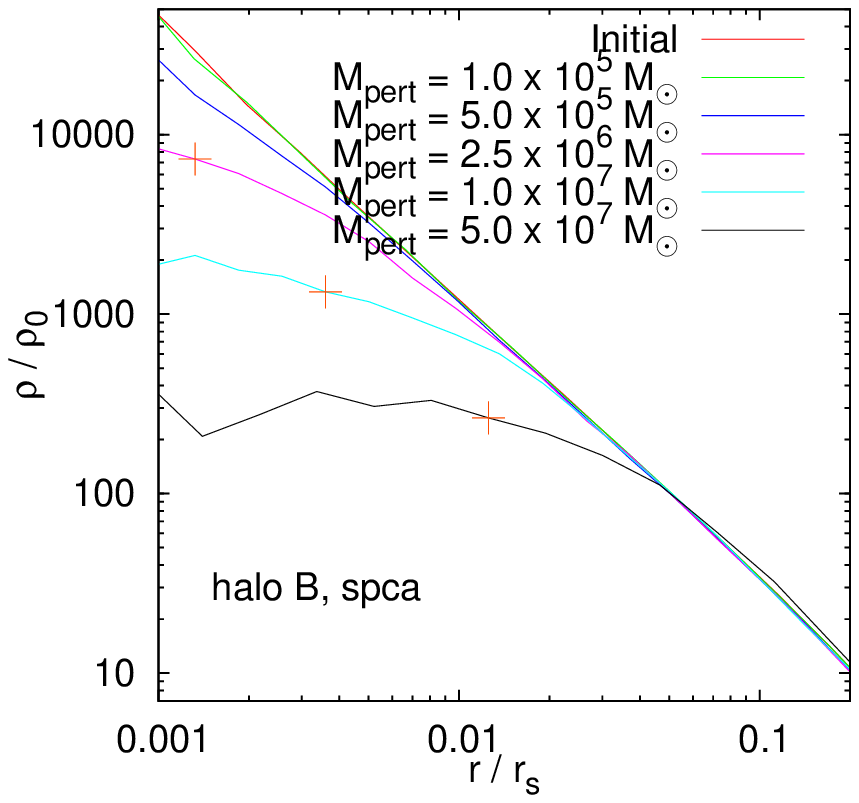}
\includegraphics[width=5.95cm]{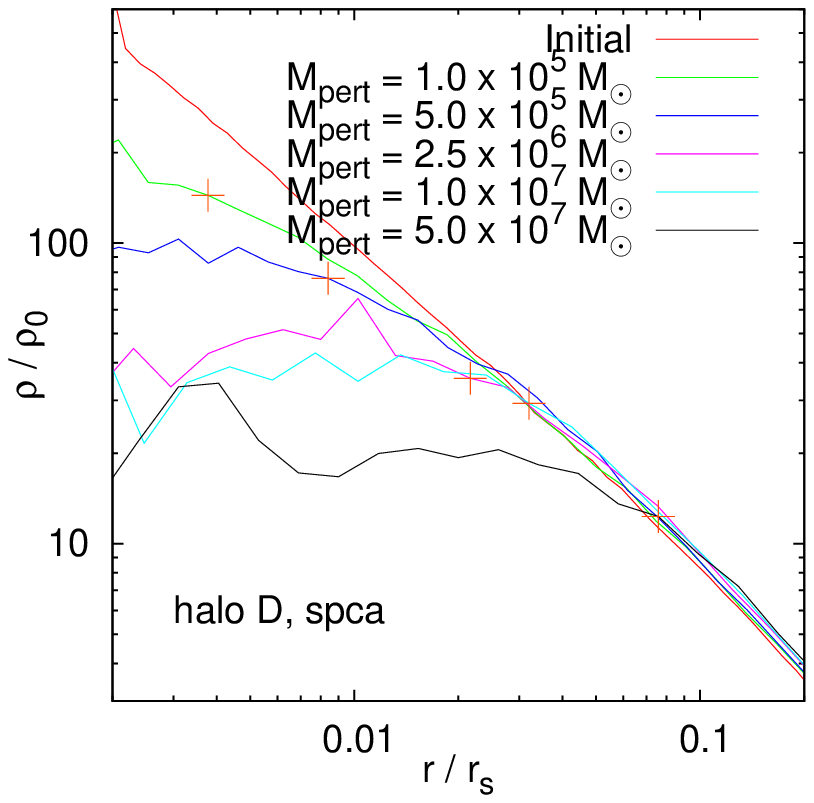}
\includegraphics[width=5.95cm]{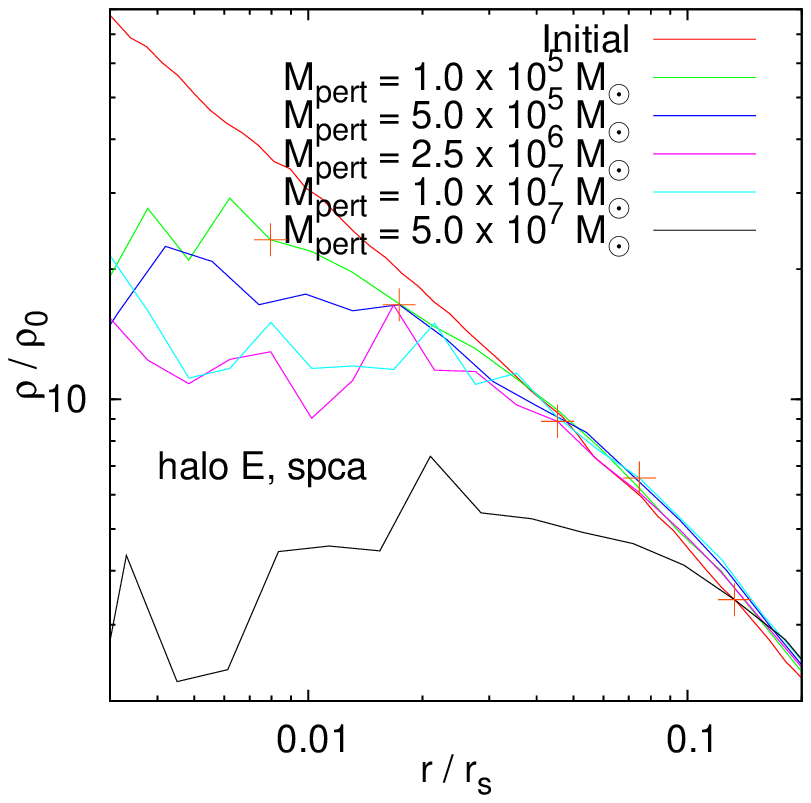}
\end{center}
\caption{Density profiles of the host halo at fpca ({\it upper panel}) and spca
({\it lower panel}) for the different nuclei masses $M_{\rm pert}$. From left
to right the panels show the haloes B, D and E. Comparing the upper and lower
panels, notice that the halo rapidly reaches a new equilibrium -- the cored
state -- between fpca and spca. The orange crosses mark the `stalling radii'
where the perturber no longer sinks via dynamical friction. Notice that these
lie at the edge of the freshly created core.}
\label{fig:denpro}
\end{figure*}

The density profiles of the respective host halo at fpca and at spca are
plotted in Figure \ref{fig:denpro}. One can clearly see that the density
distribution changes significantly from cuspy to having a core: larger 
perturber masses lead to larger constant density central regions. The orange
crosses in Figure \ref{fig:denpro} mark the `stalling radii' where the
perturber no longer sinks via dynamical friction. Notice that these lie at the
edge of the freshly created core. For this reason, from here on we define the
core radius to be the radius where the perturber stalls: $r_\mathrm{core} =
r_\mathrm{stall}$. 

\subsubsection{Why do cusps become cores?} 

\begin{figure*}
\begin{center}
\includegraphics[width=4.162cm]{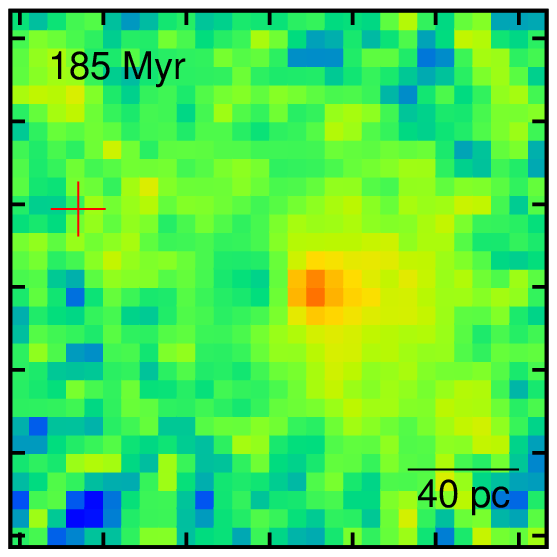}
\includegraphics[width=4.162cm]{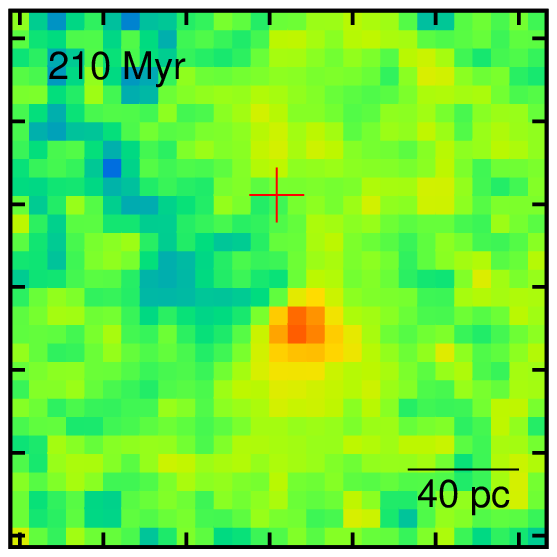}
\includegraphics[width=4.162cm]{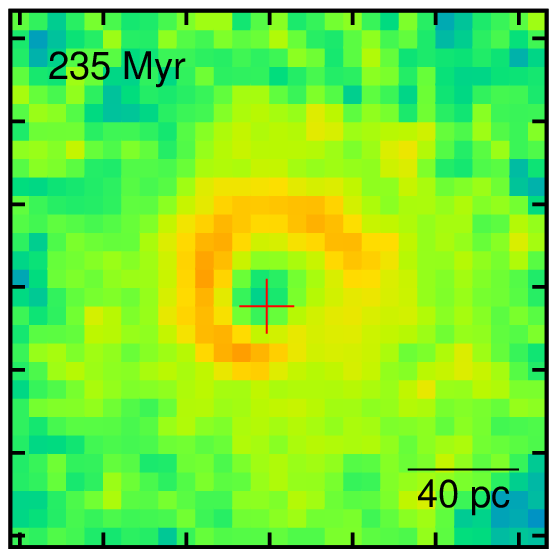}
\includegraphics[width=5.275cm]{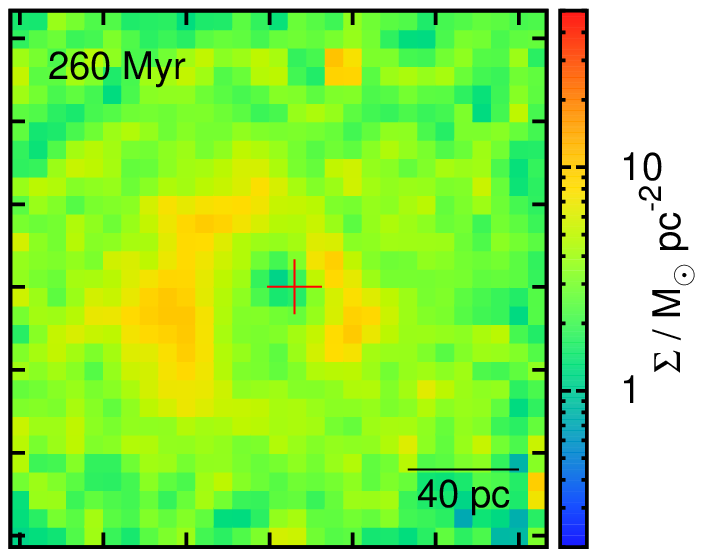}
\includegraphics[width=4.162cm]{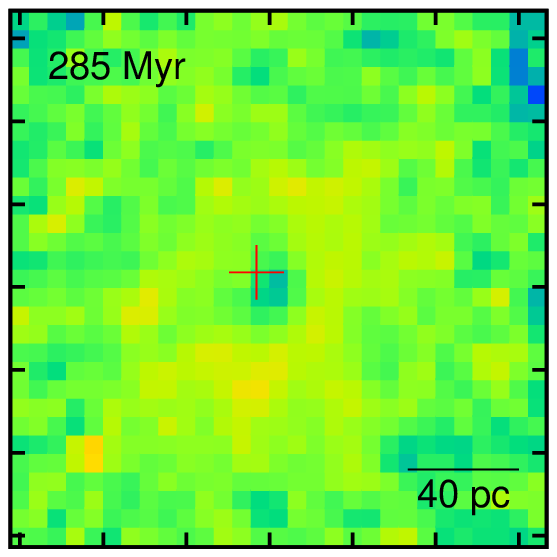}
\includegraphics[width=4.162cm]{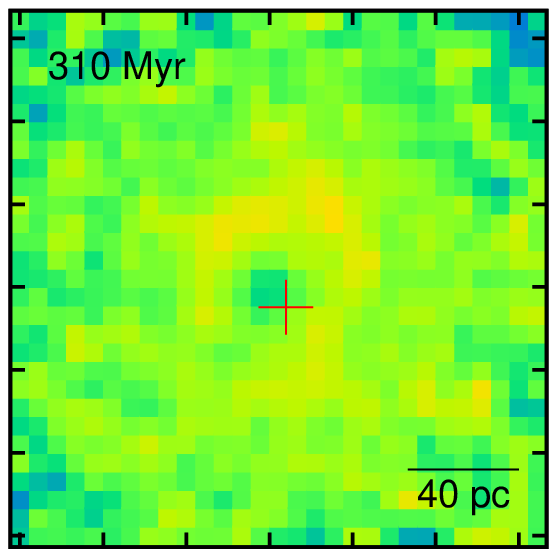}
\includegraphics[width=4.162cm]{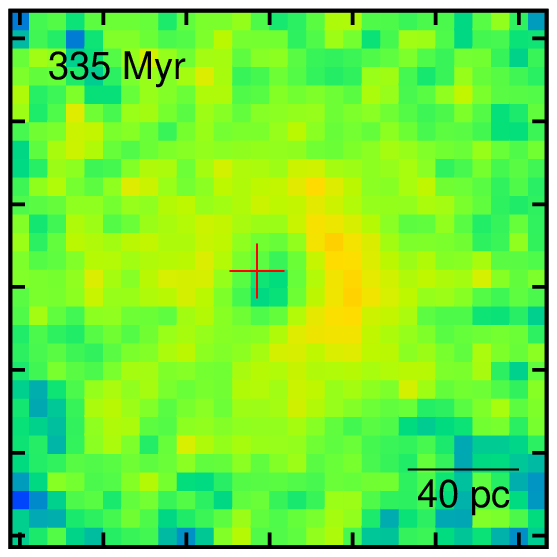}
\includegraphics[width=5.275cm]{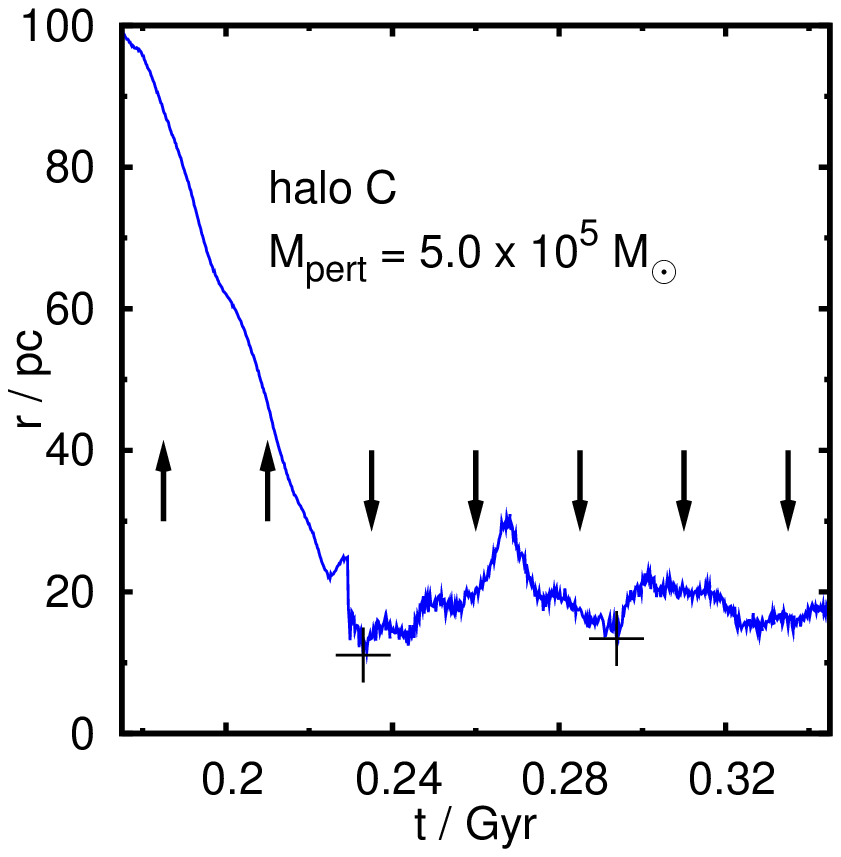}
\end{center}
\caption{194 pc $\times$ 194 pc density contour maps of background particles
within halo C. We select only particles whose orbit lies within the plane of
the perturber. The perturber mass was $5 \times$ 10$^5$\,M$_\odot$ and orbits
{\it anticlockwise} on this figure (see red crosses). The different panels show
output times:  185\,Myr, 210\,Myr, 235\,Myr (the time of the fpca), 260\,Myr,
285\,Myr, 310\,Myr and 335\,Myr. The graph in the lower right corner shows the
trajectory from Figure \ref{fig:rgc0}. The arrows indicate the above output
times and the crosses mark the first and second points of closest approach
(fpca and spca).}
\label{fig:justina}
\end{figure*}

Figure \ref{fig:justina} shows 194 pc $\times$ 194 pc density contour maps of
background particles within halo C. We select only particles whose orbit lies
within the plane of the perturber. The perturber mass was $5 \times$
10$^5$\,M$_\odot$ and orbits {\it anticlockwise} on this figure (see red
crosses). The different panels show output times: 185\,Myr, 210\,Myr, 235\,Myr
(the time of the fpca), 260\,Myr, 285\,Myr, 310\,Myr and 335\,Myr. The graph in
the lower right corner shows the trajectory from Figure \ref{fig:rgc0}. The
arrows indicate the above output times and the crosses mark the first and
second points of closest approach (fpca and spca).

Notice that the cusp in halo C becomes tidally torn into an arc at $\sim 235$
Myr. This suggests that tidal shredding of the cusp by the perturber is
responsible for the cusp-core transformation. If true, then we would expect
cusp-core transformations to occur at approximately the tidal radius $r_{\rm
t}$. This is where the mean density of the perturber roughly matches the mean
density of the background $\bar{\rho}_{\rm pert} (r_{\rm t}) \sim
\bar{\rho}_{\rm back}(r_{\rm t})$, or equivalently where the mass of the
perturber roughly matches the enclosed mass of the background $M_{\rm pert}
\sim M_{\rm back}(r_{\rm t})$ \citep{justin2}. 

\begin{figure}
\includegraphics[width=8.63cm]{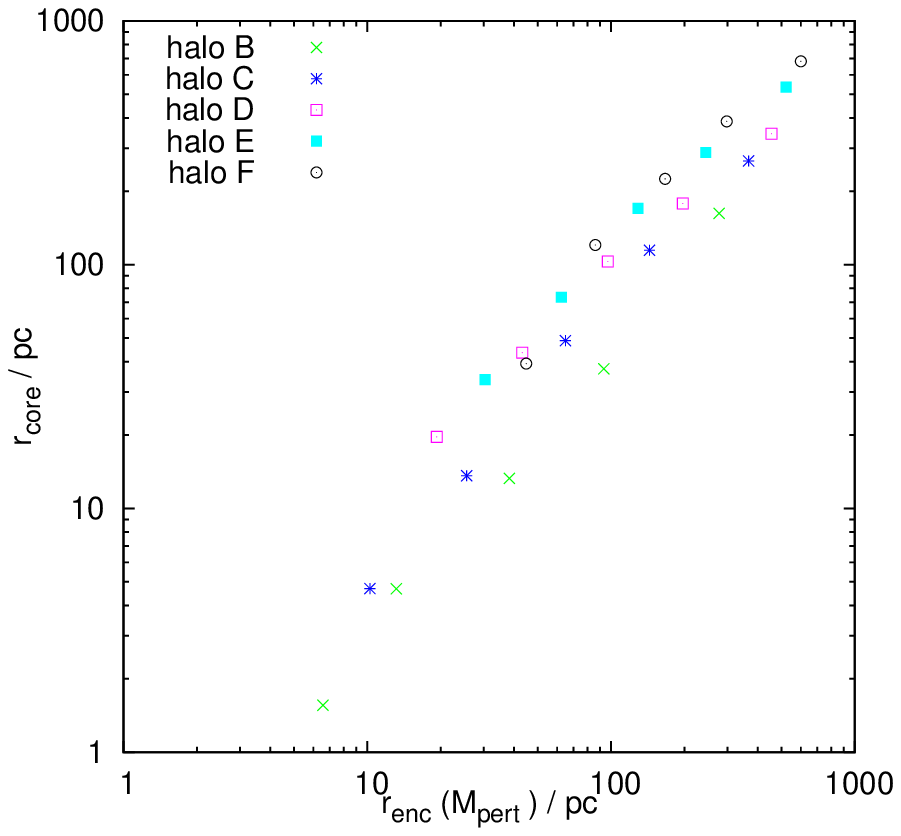}
\caption{The core radius as a function of the radius which encloses the mass
of the perturber in the initial conditions. The relation is linear and
indicates that the cusp is disrupted to a radius that contains a similar mass
as the sinking perturber.}
\label{fig:ueb}
\end{figure}

\begin{figure*}
\begin{center}
\includegraphics[width=8.82cm]{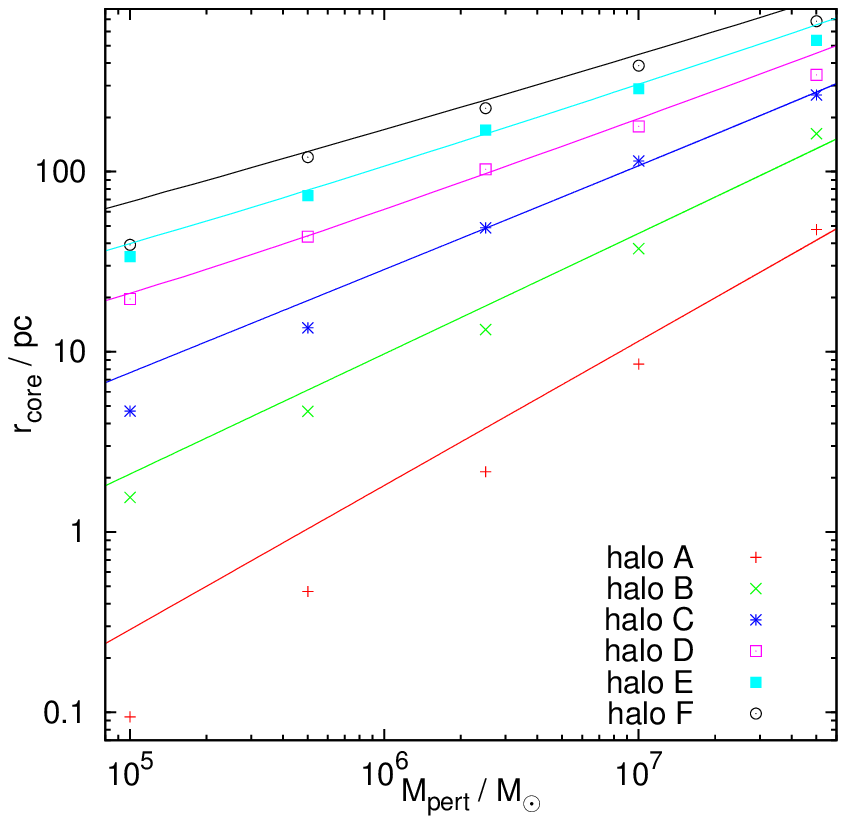}
\includegraphics[width=9.12cm]{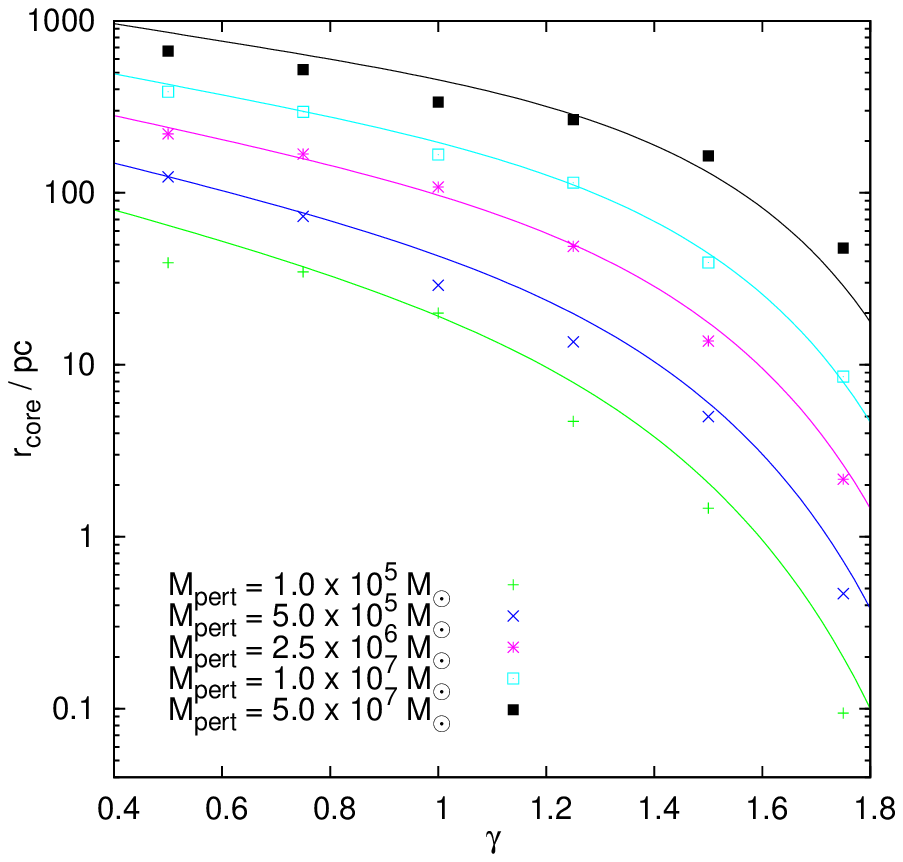}
\end{center}
\caption{The core radii as a function of $M_{\rm pert}$ for different values of
$\gamma$ ({\it left}) and as a function of $\gamma$ for different
values of $M_{\rm pert}$ ({\it right}). Our model predictions from equation
(\ref{eq:hash}) are also shown overlaid (solid lines).}
\label{fig:f13}
\end{figure*}

Figure \ref{fig:ueb} shows the core radius $r_{\rm core}$ as a function of the
radius which encloses the mass of the perturber in the initial conditions. We
find a near-linear correlation, as expected from the above hypothesis. More
concretely, we can write:

\begin{equation}
M_{\rm pert} \sim M_{\rm back}(r_{\rm t})
\end{equation}
And from equation (\ref{eq:den}), assuming that $r_{\rm t} \ll r_{\rm s}$, this
gives: 
\begin{equation}
{r_{\rm t} \over r_{\rm s}} \sim \left[\frac{(3 - \gamma) M_{\rm pert}}{4 \pi
\rho_0 r_{\rm s}^3}\right]^{\frac{1}{3-\gamma}}
\label{eq:star}
\end{equation}
We can then relate this tidal radius to the core radius via some factor
$f(\gamma)$: 
\begin{equation}
{r_{\rm core} \over r_{\rm s}} = f(\gamma) \left[\frac{(3 - \gamma)
M_{\rm pert}}{4 \pi \rho_0 r_{\rm s}^3}\right]^{\frac{1}{3-\gamma}}
\label{eq:hash}
\end{equation}
We find $f(\gamma) = 2 - \gamma$ gives an excellent fit to our simulation
results (see Figure \ref{fig:f13}).

\begin{figure}
\includegraphics[width=8.63cm]{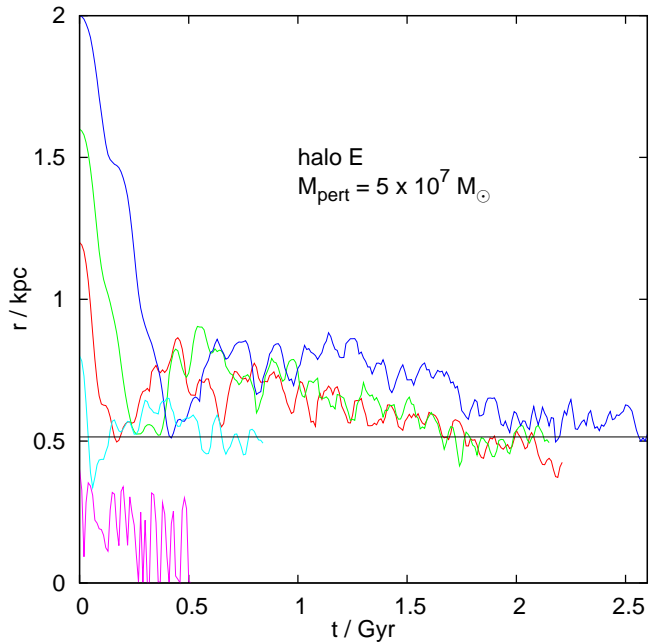}
\caption{Simulated position of a single perturber with identical mass ($5
\times 10^7$ M$_\odot$) but different initial radii within halo E as a function
of time. The core radius is indicated by the horizontal black line.}
\label{fig:f20}
\end{figure}

In Figure \ref{fig:f20}, we show the core radius $r_{\rm core}$ as a function of
the initial radius $r_{\rm i}$. If tidal shredding is responsible for cusp-core
transformations, then there should be no dependence of $r_{\rm core}$ on
$r_{\rm i}$. This is indeed the case as long as $r_{\rm i} > r_{\rm core}$.
(Note that since all of the simulations had to be started outside of $r_{\rm
core}$, the very heavy perturbers had to be started further out than the light
perturbers.)

Our analytic and numerical results can be compared with previous studies in the
literature. \citet{amr} use a semi-analytical Monte Carlo approach based on the
Chandrasekhar approximation to estimate core sizes. They let $100 - 500$
perturbers, which have a combined mass of 10\% of the host halo, sink into a
$\gamma = 1.0$ halo. They find $r_{\rm core} \approx r_{\rm s}$, where they
define  $r_{\rm core}$ as the region within which $\rho \approx \rho_0$ is
valid. Putting $M_{\rm pert} / M_{\rm vir} = 0.1$ into equation (\ref{eq:hash})
we find $r_{\rm core} = 0.57\, r_{\rm s}$. \citet{kaufmaennchen} let 10 live
globular clusters, which have a combined mass of 0.28\% of the host halo, sink
into a $\gamma = 1.0$ halo. They do not measure the core size explicitly, but
in their Figure 1 the final dark line becomes very shallow around 50\,pc.
Putting $M_{\rm pert} / M_{\rm vir} = 0.0028$ into equation (\ref{eq:hash}) we
find $r_{\rm core} = 40$ pc. Thus, our estimates agree very well with what has
already been published in the literature.

\subsubsection{Core-stalling} 

Once a core is formed, the perturbers stall and do not sink over many dynamical
times (see Figure \ref{fig:rgc0}). The mechanism for core stalling was
discussed in detail in \cite{justin}. They suggested that the perturber
and background find an equilibrium state where the background moves on
epicycles around the perturber, leading to no net momentum transfer. Recently,
\citet{inoue2} have also found evidence for such a state in their simulations.
A prediction of this is that the background has an over-density in the plane
of the perturber that lags the motion of the perturber \citep[see][their
Figure 4]{justin}. This can be seen in Figure \ref{fig:justina} at late times
(recall that the perturber moves in an anticlockwise fashion in this figure).

\section{Applications}

\subsection{Dark matter annihilation}
\begin{figure}
\includegraphics[width=8.63cm]{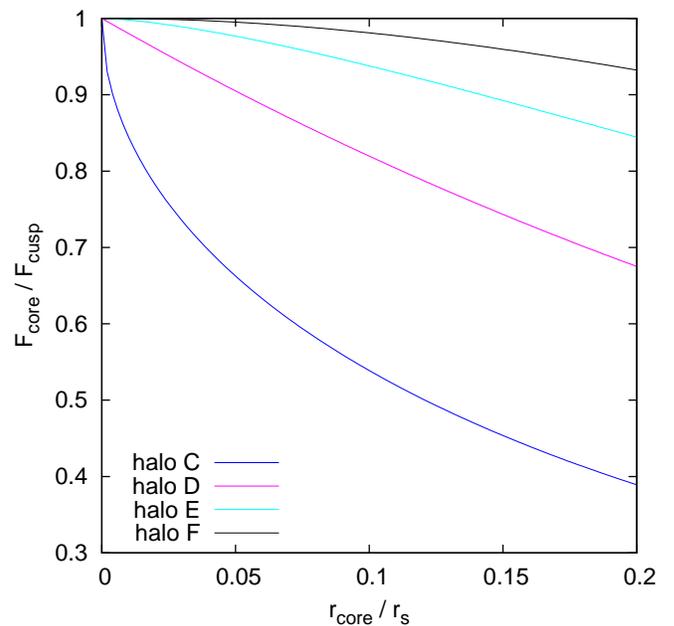}
\caption{The flux of annihilation products from different host haloes after the
central cusp core transformation relative to the untransformed initial cusp.
Typical core sizes ($r_{\rm c} \approx 0.04 r_{\rm s}$) can lead to a decrease
in flux of up to one third. Haloes A and B are not shown because the
corresponding values for inner log density slopes $\gamma \ge 1.5$ diverge.}
\label{fig:flux}
\end{figure}

The change from cuspy to cored central density is important for the expected
annihilation signal from a weakly interactive massive (WIMP) dark matter
particle, since the annihilation signal goes as the density squared. The net
flux coming from WIMP annihilation is given by \citep[e.g,][]{savvas,oleg}:
\begin{equation}
F = k \int^{\infty}_{\rm r_{min}}{4 \pi r^2 \rho(r)^2 {\rm d} r}
\label{eq:annihil}
\end{equation}
\noindent
where the dependence of the flux on the WIMP mass and interaction cross section
is wrapped up inside the constant $k$. The lower bound $r_{\rm min}$ is defined
as the central region of the host halo in which the neutralinos have already
annihilated. The required number density for this to happen can be estimated
using:
\begin{equation}
t_{\rm h} = {1 \over n \sigma v}
\label{eq:lowbound}
\end{equation}
where $t_{\rm h} \approx 13$ Gyr is the Hubble time, $\sigma v \approx 10^{-30}$
cm$^3$s$^{-1}$ is a typical cross section and $n$ is the number density of
neutralinos. For more details see \citet{calcaneo}. The minimum radius can now
be computed (for a sinking perturber of a given mass) by  comparing this
minimum number density with the density profile of the lower panel in Figure
\ref{fig:denpro}. Assuming a WIMP mass of 100 GeV and deploying the above
mentioned density profile, $r_{\rm min}$ is of order $10^{-14}$ pc. Figure
\ref{fig:flux} shows the resulting annihilation flux. It is more or less
independent of the assumed WIMP mass, because this mass only goes into the
calculation of $r_{\rm min}$, which is very small. For typical core sizes of
$r_{\rm core}\simeq 0.04\,r_{\rm s}$, core creation can lead to a decrease in
flux of up to a third. A much weaker effect would result from a single sinking
star; for example a 10\,M$_\odot$ star would create a core of radius 0.34\,pc in
our fiducial halo. However, note that the other core-formation mechanisms
discussed in section \ref{sec:intro} may play a more important role than the
dynamical friction mechanism discussed here. In this case, the annihilation
signal could be reduced even further. An increase in the expected signal can
also occur if dark matter adiabatically contracts due to gas cooling at the
centre of dark matter halos \citep[e.g.][]{young, blumenthal}. Such
astrophysical uncertainties make it currently challenging to predict the
expected annihilation signal for a given dark matter model.

\subsection{A new model for close binary nuclei -- the `stalled binary' model} 

There are a number of systems in the universe which show evidence for close
binary nuclei \citep[with projected separation $< 100$\,pc; e.g.][]{lauer1,
lauer, lauer2, bender, houghton, mast, victor}. Only three of these are
unambiguous binary nuclei systems -- M31, NGC~4486B and VCC~128; the rest are
more poorly resolved systems that show strongly asymmetric central light
distributions.

The standard model for these close `binaries' has become the \cite{tremaine}
eccentric disc model originally proposed to explain M31 \citep{lauer2}. In this
scenario, the two central peaks are really an artifact of a central eccentric
disc of stars orbiting a black hole. The fainter peak is associated with stars
orbiting at pericentre in the disc. The brighter peak is associated with stars
at apocentre where they linger for long times. Key to the success of this model
is a near-Keplerian central potential. This ensures that there is no orbital
precession so that the bunching at apocentre is maintained over many orbital
periods. 

For M31, the eccentric disc model works remarkably well since its binary
`nuclei' have a projected separation smaller than the sphere of influence of
the central black hole \citep[$\sim 1.8$\,pc; ][]{peiris, bender}. In addition,
more recent observations suggest that the M31 nucleus is really a triple
system, where the third nucleus P3 is centred on the central black hole
\citep{bender}. 

While the eccentric disc model has been a success for M31, it is not clear that
it works well for the other `close binary' systems observed to date. Many have
significantly larger projected separations than M31 \citep{lauer, lauer2,
victor}, which means that their central potentials may deviate from being pure
Keplerian. In addition, both VCC~128 and NGC~4486B have binary nuclei with very
similar magnitudes. This requires a tilted ring model, rather than an eccentric
disc of stars \citep{lauer, victor}, which makes the `eccentric disc' model
seem less attractive. 

When the double nucleus of M31 was first discovered, \cite{lauer1} speculated
that the two bright nuclei really were just that -- one from M31 and the other
the cannibalised centre of a smaller merged galaxy. The primary argument
against this was that dynamical friction would cause the nuclei to rapidly
coalesce. However, as we have shown here, as binary nuclei sink via dynamical
friction, they create a central constant density core. They then stall at the
edge of this core experiencing no further friction over many dynamical times.
This suggests a new model for the formation of close binary nuclei -- the
`stalled binary' model. 

A particularly interesting candidate system for a `stalled binary' is VCC~128.
This has two nuclei that have similar brightness, and a projected separation of
$\sim 32$\,pc \citep{victor} which presents some tension for the eccentric disc
model. As we shall show next, however, the `stalled binary' model works
remarkably well. VCC~128 is also especially interesting since it appears to be
dark matter dominated at all radii. As a result -- if its binary nucleus is a
`stalled binary' -- VCC~128 gives us a unique opportunity to constrain the
central log-slope of the dark matter density profile on very small scales. We
consider this special case next. 

\subsubsection{The binary nuclei in VCC~128}
VCC~128 is a dwarf spheroidal galaxy (dSph) at the outskirts of the Virgo
cluster with a very close binary nucleus \citep{victor}. The two nuclei are
similar in their appearance with masses estimated to be around $5 \times 10^5$
M$_\odot$. The projected distance of the two nuclei in VCC~128 is 32\,pc.
\citet{victor} suggest, because the two nuclei have very similar colours and
magnitudes that this could be evidence for a nuclear disc around a supermassive
black hole (SMBH), a situation as in NGC~4486B \citep{lauer} and similar to the
one in M31 \citep{tremaine}. However, it is not confirmed that such an SMBH can
exist in a dwarf galaxy like VCC~128. \citet{buyle} do not find statistically
significant evidence for a stellar disk orbiting a central massive black hole
from their recent radio continuum observations of VCC~128. \citet{ferrarese}
found from a sample of 36 galaxies, tentative evidence that SMBH formation
becomes inefficient in haloes below a dynamical mass of $\sim 5\times 10^{11}$
M$_\odot$, though more recent work may suggest otherwise \citep{ferrarese6,
wehner}. As such, it is interesting to consider whether our new `stalled binary'
model can explain VCC~128's binary nucleus.

The dwarf galaxy VCC~128 is likely dark matter dominated at all radii. We
estimate its stellar mass distribution in two different ways. Firstly, we use
the \citet{sersic1,sersic2} profile parameters derived in \cite{victor}
normalised to give a total luminosity in the B-band of M$_{\rm B} = -15.5$\,mag:

\begin{equation}
\Sigma(R) = \Upsilon_{\rm B} I_{\rm e} \exp\left\{-b_{\rm n}\left[\left(
\frac{R}{R_{\rm e}}\right)^{\frac{1}{\rm n}}-1\right]\right\}
\end{equation}
with $b_{\rm n} = 1.9992 n - 0.3271$ \citep{simon}, $n = 0.55$, $R_{\rm e} =
14.5$\,arcsec and assuming a B-band mass to light ratio $\Upsilon_{\rm B} = 3$
for dE galaxies as in \cite{readtrent}. We then de-project the stellar mass
distribution using the usual Abel integral equation (that assumes spherical
symmetry):

\begin{equation}
\rho(r) = -\frac{1}{\pi}\int_{\rm r}^{\infty} \frac{{\rm d}\Sigma(R)}{{\rm
d}R}\frac{{\rm d}R}{\sqrt{R^2 - r^2}}.
\end{equation}

The resulting density and cumulative mass distributions are given by the solid
lines in Figure \ref{fig:f1}. Secondly, we use the stellar masses derived from
fits to the spectral energy distribution (SED) of the galaxy nucleus and the
whole galaxy as given in \cite{victor}. These are overlaid on the right panel
of Figure \ref{fig:f1} ({\it crosses}) and give an excellent match to the
cumulative mass distribution derived from the S\'ersic fit to the light
profile. We assume from here on that the dark matter is dynamically dominant
and that the stars are to a good approximation a massless tracer population. 

\begin{figure}
\includegraphics[width=8.63cm]{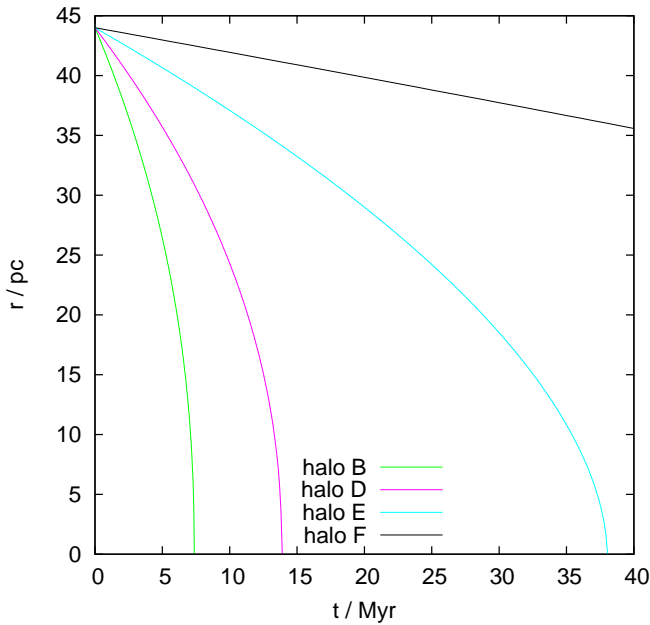}
\caption{Analytical computed position as a function of time, during the
Chandrasekhar sink-in period for four of our haloes, assuming an initial true
distance of the nucleus from the centre of 44 pc. Haloes A and C are omited.}
\label{fig:f3}
\end{figure}

Figure \ref{fig:f3} shows the time taken for two nuclei that have masses $5
\times 10^5$\,M$_\odot$ and initial separation of 44\,pc to sink via dynamical
friction in halos B, D, E and F, assuming that equation (\ref{eq:L1}) fully
describes the friction process. We chose this initial separation of 44\,pc
since this is the most likely deprojected distance of the two nuclei (their
observed separation on the sky is 32\,pc). In all haloes with cusp slopes
steeper than 0.75, the nuclei coalesce rapidly. This suggests that, were
equation (\ref{eq:L1}) the whole story, it would be very unlikely to observe
such a close double nucleus in any dwarf galaxy in Virgo. However, as we have
demonstrated in the previous section, the merging nuclei will create a small
core and stall indefinitely, leading to a much higher probability of observing
double nuclei.

\begin{figure}
\includegraphics[width=8.63cm]{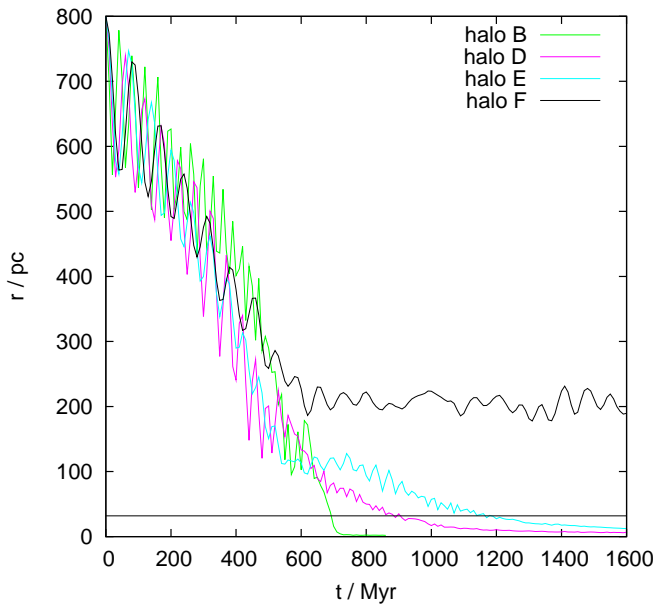}
\caption{Separation of the two perturbers as a function of time, assuming an
initial true separation of 800\,pc and perpendicular orbits. The horizontal
black line shows the observed projected distances in VCC~128. Haloes A and C
are omited.}
\label{fig:f5}
\end{figure}

Therefore we ran an additional series of simulations with two perturbers on
either coplanar or perpendicular orbits. Again we used the haloes presented in
Table \ref{tab:haloes}. The stalling behaviour is shown in Figure \ref{fig:f5}.
In haloes B, D and E, the nuclei do not stall. Only in halo F do they stall
above 32\,pc. These results suggest for slopes of 1.0 or steeper, this stalling
mechanism does not crucially change the results we derived analytically in this
section. However for slopes shallower than 1.0 it affects these results quite
dramatically. It is interesting that the most recent CDM halo simulations 
have cusps shallower than $0.7$ on these scales \citep{ghalo, vialactea,
aquarius}.

To summarise our findings: assuming that VCC~128 has a core-stalled binary
nucleus, we can exclude all steep inner log density slopes of 0.75 or higher.
Our current best explanation for VCC~128's observed binary nucleus is that the
two nuclei transformed an initially steeper density profile with an inner log
density slope between 0.75 and 0.5 into a core during its initial sinking
period. The nuclei then stalled at the edge of this freshly created core. Note
that a very shallow inner log density slope like 0.5 is also disfavoured. This
is because in this case, the nuclei stall at $\sim 100-200$\,pc (see Figure
\ref{fig:f5}) creating some tension with the observed projected separation.
Therefore we conclude that an underlying dark matter halo with $\gamma \sim
0.5 - 0.75$ at $\sim 1$\% of the virial radius provides the best explanation
for the observations. 

One should note here that this cusp-core transformation mechanism does not
explain the dynamical friction timescale problem of Fornax within a
$\Lambda$CDM ($\gamma =$ 1.0) halo \citep[see Figure 3 of][]{mich}, because in
that case the globular clusters in Fornax are too light and too distant from
the centre of their host galaxy for this cusp-core transformation mechanism to
play a significant role. Equation (\ref{eq:hash}) predicts for Fornax (assuming
$\gamma = 1.0$) a core of only 44\,pc. This is far smaller then the projected
radius of the innermost globular cluster in Fornax, which is at 240\,pc.

\section{Conclusions}
We have performed a detailed investigation into the disruption of central cusps
via the transfer of energy from sinking massive objects. Constant density inner
regions form at the radius where the enclosed mass approximately matches the
mass of the infalling body. We explored parameter space using numerical
simulations and gave an empirical relation for the size of the resulting core
within structures that have different initial cusp slopes. We went on to
demonstrate that infalling bodies always stall at the edge of these newly
formed cores, experiencing no dynamical friction over many dynamical times. 

As applications, we considered the resulting decrease in the WIMP annihilation
flux due to centrally destroyed cusps; and we presented a new theory for the
formation of close binary nuclei -- the `stalled binary' model. Our key results
are as follows: 

\begin{enumerate}
\item Core formation due to sinking massive objects can soften a central dark
matter cusp reducing the expected WIMP annihilation flux (predicted by
structure formation simulations that model dark matter in the absence of
baryons) by up to a third. 

\item Core formation due to sinking massive objects could help to alleviate the
long-standing cusp-core problem \citep[see eg:][]{spekkens,moore2,deblok}. From
equation (\ref{eq:hash}), a $\sim$1\,kpc sized core will form from perturber
having $\sim$1\% of the mass of the host. This recovers the earlier
results of \citet{amr} and \cite{jardel}. However, such massive infalling
perturbers must later disrupt or be removed in order to be consistent with the
low surface density of stars and gas observed in galaxies where the cusp-core
problem is most apparent.

\item Infalling nuclei at the centres of galaxies will evacuate a core and
stall indefinitely, provided that the initial background density is not
significantly steeper than $r^{-1}$. This could explain a number of binary
nuclei systems in the Universe.

\item We focused on the special binary nucleus system VCC~128 since it is dark
matter dominated at all radii. Assuming that its binary nucleus can be
explained by our `stalled binary' model, we found that the initial inner log
density slope $\gamma$ of the dark matter halo of VCC~128 must be $0.5 < \gamma
< 0.75$ at $\sim 0.1\%$ of the virial radius. For $\gamma > 0.75$ initially, the
dynamical friction sink-in time is so small in comparison to the lifetime of
the galaxy that we run into a fine tuning problem. For $\gamma < 0.5$
initially, the nuclei stall far beyond their current projected separation of
32\,pc. For $0.5 < \gamma < 0.75$ initially, the nuclei create a central
constant density core of separation $\sim 40$\,pc after which they stall
indefinitely. Our preferred inner slopes are consistent with those found in the
recent billion particle CDM halo simulations of \citet{ghalo},
\citet{vialactea} and \citet{aquarius}.
\end{enumerate}

\acknowledgments
It is a pleasure to thank Ioannis Sideris and Eyal Neistein for carefully
reading the manuscript. For all $N$-body simulations we used \textsc{Pkdgrav2}
\citep{stadel}, a multi-stepping tree code developed by Joachim Stadel. All
computations were made on the zBox2 supercomputer at the University of
Z\"urich. Special thanks go to Doug Potter for bringing it to life. Tobias
Goerdt is a Minerva fellow.

\end{document}